\documentclass{llncs}

\usepackage{graphicx}
\usepackage{cite}
\usepackage{picinpar}
\usepackage{amsmath}
\usepackage{stfloats}
\usepackage{url}
\usepackage{subfigure}
\usepackage{flushend}
\usepackage[utf8]{inputenc}
\usepackage{colortbl}
\usepackage{soul}
\usepackage{multirow}
\usepackage{pifont}
\usepackage{color}
\usepackage{alltt}
\usepackage[hidelinks]{hyperref}
\usepackage{enumerate}
\usepackage{siunitx}
\usepackage{breakurl}
\usepackage{epstopdf}
\usepackage{pbox}
\usepackage{cuted}
\usepackage{mathtools}
\usepackage{mathpazo}
\usepackage{color}
\usepackage{wrapfig}
\usepackage{}

\usepackage[T1]{fontenc}
\usepackage[table]{xcolor}
\usepackage{collcell}
\usepackage{hhline}
\usepackage{pgf}
\usepackage{multirow}

\usepackage{colortbl}
\usepackage{hhline}

\usepackage{amssymb}

\usepackage{graphicx}
\usepackage{float}
\makeatletter
\newcommand\fs@myRoundBox{\def\@fs@cfont{\bfseries}\let\@fs@capt\floatc@plain
	\def\@fs@pre{\begin{mdframed}[style=myFigureBoxStyle]}%
		\def\@fs@mid{\vspace{\abovecaptionskip}}%
		\def\@fs@post{\end{mdframed}}\let\@fs@iftopcapt\iffalse}
\makeatother

\DeclarePairedDelimiter\floor{\lfloor}{\rfloor}

\begin{document}
	
	\title{Phase transitions in swarm optimization algorithms}

	
	\author{Tom\'{a}\v{s} Vantuch\inst{1}, Ivan Zelinka\inst{1}, Andrew Adamatzky\inst{2} and Norbert Marwan\inst{3}}
	
	\institute{ 
		Department of Computer science, Technical University of Ostrava,  Czech Republic
		\and 
		Unconventional Computing Lab, UWE, Bristol, UK
		\and
		Potsdam Institute for Climate Impact Research (PIK), Transdisciplinary Concepts and Methods, Potsdam, Germany}
	
	\maketitle
	
	\begin{abstract}
		
Natural systems often exhibit chaotic behavior in their space-time evolution. Systems transiting between chaos and order manifest a potential to compute, as shown with cellular automata and artificial neural networks. We demonstrate that swarms optimisation algorithms also exhibit transitions from chaos, analogous to motion of gas molecules, when particles explore solution space disorderly, to order, when particles follow a leader, similar to molecules propagating along diffusion gradients in liquid solutions of reagents. We analyse these `phase-like' transitions in swarm optimization algorithms using recurrence quantification analysis and Lempel-Ziv complexity estimation. We demonstrate that converging and non-converging iterations of the optimization algorithms are statistically different in a view of applied chaos, complexity and predictability estimating indicators. 

		
	\end{abstract}
	
    
	\section{Introduction}

	Natural systems not rarely undergo phase transition when performing a computation (as interpreted by humans), e.g. reaction-diffusion chemical systems produce a solid precipitate representing geometrical structures~\cite{costello2017calculating}, slime mould transits from a disorderly network of `random scouting' to a prolonged filaments of protoplasmic tube connecting source of nutrients~\cite{adamatzky2016advances},  `hot ice' computer crystallizes~\cite{adamatzky2009hot}. Computation at the phase transition between chaos and order was firstly studied by Crutchfield and Young~\cite{crutchfield1988computation}, who proposed measures of complexity characterising the transition. The ideas were applied to cellular automata by Langton~\cite{langton1990computation}: a computation at the edge of chaos occurs due to gliders.  Phase transitions were also demonstrated for a genetic algorithm which fall into a chaotic regime for some initial conditions~\cite{mitchell1993revisiting,wright2001cyclic} and network traffic models~\cite{ohira1998phase}.

	Algorithmic models of evolutionary based optimization, AI and ALife possess comparable features of the systems with a higher complexity, they simulate \cite{zenil2017algorithmic,detrain2006self}. We focus  on the behavioral modes: the presence of a random or pseudo-random cycling (analogous to gaseous phase state), ordered or a stable states (analogous to solid state), or the chaotic oscillations (transitive states). Each of the modes could imply different level of a computational complexity or an algorithm performance as it was revealed on different algorithms \cite{boedecker2012information,bertschinger2004real,kadmon2015transition}. By detecting such modes we can control and dynamically tune performance of the computational systems. 
	

	A swarm-like behavior has been extensively examined in studies of Zelinka et al. \cite{zelinka2017novel} where the changing dynamics of an observed algorithm was modeled by a network structure. The relevance between network features and algorithm behavior supported the control mechanism that was able to increase the algorithm performance \cite{tomaszek2016performance}. An extensive empirical review of existing swarm based algorithms has been brought by Schut \cite{schut2010model} where approaches like collective intelligence, self-organization,
	complex adaptive systems, 
	multi-agent systems, 
	swarm intelligence
	were empirically examined and confronted with their real models which reflected several criteria for development and verification.
	
	We aim to evaluate the dynamics of optimization algorithms, inspired by evolution and swarm-like behavior. We evaluate the dynamical modes of algorithms based on predictability, complexity and chaos features. At the end, we statistically examine the difference between estimated modes, they possessed. In case of successful detection of statistically different modes and their transitions during the optimization process, the edge of chaos may be examined as well as controlling tools may be designed.
	
	\section{Theoretical background}

	\subsection{Swarm based optimization}
	
	The optimization algorithms examined in our study are representatives of bio-inspired single-objective optimization algorithms. They iteratively maintain the population of candidates migrating through the searched space. Their current position represents the solution vector $X$ of the optimized problem.
	
	
	\emph{Particle Swarm Optimization} implies that 
	the combined particle's aim towards the global leader and its previous best position~\cite{kennedy1995particle}. The composition of these two stochastically altered directions modifies its current position in order to find a better optimum of the given function. Several reviewing studies are available as extensive descriptions of the algorithm and they are also surveying proposed extensions and variations \cite{banks2007review,del2008particle}.
	
	%
	%
	%
	%
	\emph{Differential Evolution}
	(DE) was developed by R. Storn and K. Price \cite{storn1997differential} and it possesses the features of a self-organizing search as well as an evolutionary based optimization. This interconnection is deserved due to its three main stages. DE offers several strategies driving the computation of new positions for its candidates. One of them takes three random candidates to calculate an intermediate candidate which creates a new position by binary crossover with an optimized candidate $x_i$. It takes this new position only if it is better than the current one.
	
	%
	%
	%
	%
	\emph{Self-organizing migrating algorithm}
	(SOMA) is  a stochastic evolutionary algorithm was proposed by Zelinka \cite{zelinka2004soma}. Ideologically, this algorithms stands right between purely swarm optimization driven PSO and evolutionary-like DE. The entire nature of migrating individuals across the search-space is represented by steps in the defined path length and a stochastic nature of a perturbation parameter that represents specific version of the mutation.
	
	%
	%
	%
	%
	%

	\subsection{Lemplel-Ziv complexity}
	
	According to the Kolmogorov's definition of complexity, the complexity of an examined sequence $X$ is the size of a smallest binary program that produces such sequence \cite{cover2012elements}. Because this definition is way too general and any direct computation is not guaranteed within the finite time \cite{cover2012elements}, approximative techniques are frequently employed. 
	
	Lempel and Ziv designed a complexity estimation in a sense of Kolmogorov's definition, but limiting the estimated program only to two operations: recursive copy and paste \cite{lempel1976complexity}. The entire sequence based on an alphabet $\aleph$ is split into a set of unique words of unequal lengths, which is called a vocabulary. The approximated binary program making use of copy and paste operations on the vocabulary, is able to reconstruct the entire sequence. Based on the size of vocabulary ($c(X)$), the complexity is estimated as $C_{LZ}(X) = c(X)(\text{log}_kc(X) + 1)\cdot N^{-1}$, where $k$ means the size of the alphabet and $N$ is the length of the input sequence.
	%
	A natural extension for multi-dimensional LZ complexity was proposed in \cite{zozor2005lempel}. In case of a set of $l$ symbolic sequences ${X^i} (i=1,\cdots, l)$, Lempel and Ziv’s definitions remain valid if one extends the alphabet from scalar values $x_k$ to $l$-tuples elements $(x^1_k, \cdots, x^l_k)$. The joined-LZC is than calculated as $C_{LZ}(X^1,\cdots,X^l)=c(X^1,\cdots,X^l)(\text{log}_{k^2}c(X^1,\cdots,X^l) + 1)\cdot N^{-1}$.
	
	

	\subsection{Recurrence quantification analysis}

	
	The recurrence plot (RP) is the visualization of the recurrence matrix of $m$-dimen\-sional system states $\vec{x}\in \mathbb{R}^m$ \cite{marwan2007recurrence}. The closeness of these states for a given trajectory $\vec{x}_i\ (i = 1, 2,...,N)$ where $N$ is the trajectory length, is thresholded in the Heaviside step function $\Theta(\cdot)$ which results in the binary matrix of recurrence $R_{i,j} (\epsilon) = \Theta(\epsilon- \|\vec{x}_i - \vec{x}_j\|)$. The Euclidean norm is the most frequently applied distance metric $\|\cdot\|$ and the threshold value $\epsilon$ can be chosen according to several techniques \cite{koebbe1992use,zbilut2002recurrence,zbilut1992embeddings,marwan2007recurrence,schinkel2008}.
	
	If only one-dimensional time series is given, the phase space trajectory has to be reconstructed from the time series $\{u_i\}^N_{i=1}$, e.g., by using the time-delay embedding $\vec{x}_i = (u_i, u_{i+\tau} ,...,u_{i+(m-1)\tau})$, where $m$ is the embedding dimension and $\tau$ is the embedding delay \cite{packard1980geometry}. The parameters $m$ and $\tau$ may be found using methods based on false nearest neighbors and auto-correlation \cite{kantz97}.

	
	The RQA measures applied in this experiment describe the predictability and level of chaos in the observed system. Determinism is defined as the percentage of points that form diagonal lines (Eq. \ref{det_eq})
	
	\begin{equation}
	\label{det_eq}
	DET = \sum\limits_{l=2}^NlP(l)\left[\sum\limits_{l=1}^NlP(l)\right]^{-1}
	\end{equation}
	where $P(l)$  is the histogram of the lengths $l$ of the diagonal lines \cite{marwan2007recurrence}. Its values, ranging between zero and one,  estimate the predictability of the system.
	
%
	%
	Divergence is related to the sum of the positive Lyapunov exponents, naturally computing the amount of chaos in the system, and it is defined as follows
	
	\begin{equation}
	\label{div_eq}
	DIV = L_{\max}^{-1}, \qquad L_{\max} = \text{max}(\{l_i;i=1,\cdots,N_l\})
	\end{equation}
	where $L_{\max}$ is the longest diagonal line in the RP (excluding the
	main diagonal line)\cite{marwan2007recurrence}.
	

	
	\section{Experiment design}
	
	\paragraph{Data preparation.}

	
	All three examined algorithms attempted to optimize one common fitness-function, the Rastrigin function, because of its frequent application with similar manners and its dimensional scalability that satisfies our testing purposes: $f(x) = A\cdot n + \sum\limits_{i=1}^n (x_i^2 - A\cdot\text{cos}(2\pi x_i))$, where $A$=10 and $x_{i}\in [-5.12,5.12]$. The function has a global minimum at $x$ = 0 where $f(x) = 0$.

	%
	The adjustment of the optimization algorithms was tuned by random search hyper-parameter optimization \cite{bergstra2012random} in order to find the optimal adjustment to perform the best possible convergence. The only fixed hyper-parameters were the dimension of the optimized function (it also affected the dimension of the particles, $D$ = 10) and the population size of the algorithm ($NP$ = {40, 60, 100} - it varied in order to see the affect of population size on the appearing dynamics). The rest of the hyper-parameters were optimized in the ranges according to Table \ref{dynamic_hyper}. 

	\begin{table}[h]
		\caption{The value ranges of hyper-parameters of optimization algorithms to be adjusted with their meaning.}
		\label{dynamic_hyper}
		\begin{center}
			\begin{tabular}{ l | c | l | l }
				Parameter& Algorithm & Meaning & Value \\
			\hline
				$c_1$ & PSO & global best position multiplier & $\langle 0.5, 1.5 \rangle$  \\
			$c_2$ & PSO &  local best position multiplier & $\langle 0.5, 1.5 \rangle$ \\
			$w$ & PSO & inertia weight & $\langle 0.5, 0.95 \rangle$ \\
			\hline
			$F$ & DE & differential weight & $\langle 0.1, 1.0 \rangle$ \\
			$Cr$ & DE & crossover probability & $\langle 0.1, 1.0 \rangle$ \\
		\hline
		$prt$ & SOMA & pertubation probability & $\langle 0.1, 1.0 \rangle$ \\
			step size & SOMA & size of the performed step & $\langle 0.1, 1.0 \rangle$ \\
	\hline
		\end{tabular}
		\end{center}
	\end{table}
	
	The behavior of the optimization algorithms is represented by the positions ($X_{t_1} = \{x_{t_1,1}, x_{t_1,2}, \cdots, x_{t_1,D}\}$) taken by their population members ($P = {p_1,p_2,\cdots,p_N}$) during their migrations/iterations ($p_1={X_{t_1,1},X_{t_2,1},\cdots,X_{t_m,1}}$). All of them are stored for the further examination. The time windows $w$ of iterations are taken and transfered into matrices of particles positions where columns are particle's coordinates and rows are ordered particles by their population number and time 
	
	($P_{w_{i}} = \{x_{t_i,1}, x_{t_i,2}, \cdots, x_{t_i,N}, x_{t_{i+1},1}, x_{t_{i+1},2} \cdots, x_{t_{i+1},N}, \cdots x_{t_{i+w},N}\}$).
	
	\paragraph{Convergence.}
	
	Applying the before-mentioned algorithms' hyper-parameters, the optimization converged towards an optimum. In case of our experiment, the exclusive finding of a global optimum does not play such an important role as the fact that algorithms converge towards a fixed point performing various changes and interactions inside of their swarm. 
	

    \begin{figure}[!tbp]
		\centering
		\subfigure[]{\includegraphics[scale=0.7]{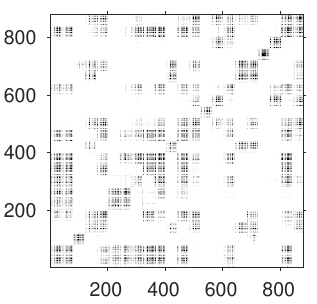}} 
		\subfigure[]{\includegraphics[scale=0.7]{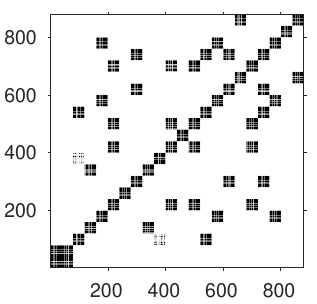}} 
		\subfigure[]{\includegraphics[scale=0.7]{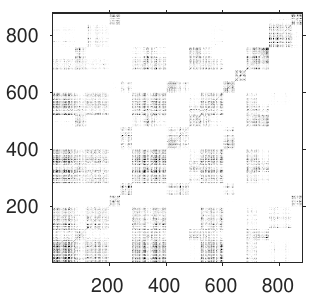}}\\
		\subfigure[]{\includegraphics[scale=0.7]{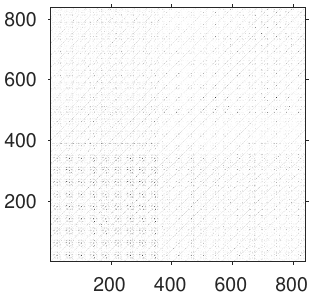}}
		\subfigure[]{\includegraphics[scale=0.7]{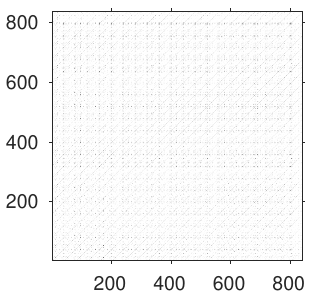}}
		\subfigure[]{\includegraphics[scale=0.7]{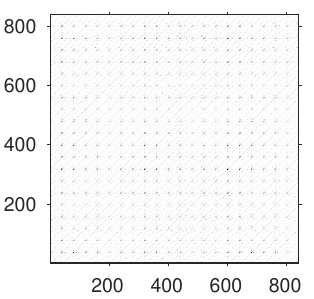}}\\
		\subfigure[]{\includegraphics[scale=0.7]{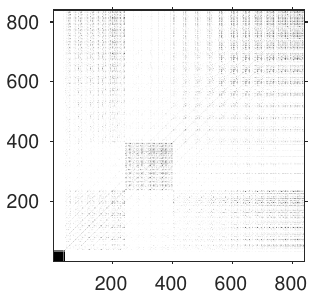}}
		\subfigure[]{\includegraphics[scale=0.7]{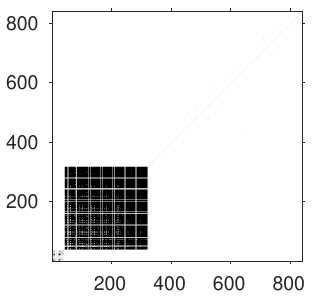}}
		\subfigure[]{\includegraphics[scale=0.7]{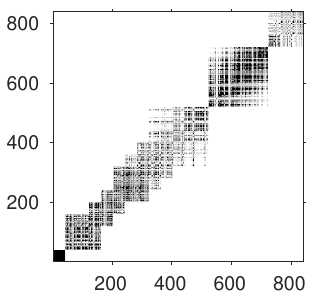}}
		\caption{Recurrence plots of the PSO (abc), DE (def), and SOMA (ghi) behavior calculated as similarities among the particles' positions $X_t$ grouped into the windows of populations $P_{w_i}$ during their (adg)~``post-initial'' (10th migration), 
        (beh)~``top-converging'' (60th migration) and (cfi)~``post-converging'' (400th migration) phase.}
		\label{pso_de_soma}
	\end{figure} 

The changes and interactions inside of their migrating populations are not usually visible in convergence plots, however changes during the convergence may be estimated using recurrence plots. For this purpose, three selected windows of algorithms' iterations were visualized to spot the differences among them. Figure~\ref{pso_de_soma} illustrates how phases of the algorithm convergences are reflected in recurrence plots.

	%
	%
	
	\paragraph{Complexity estimation.}
	The obtained matrix $P_{w_{i}}$ served as input for a joint Lempel-Ziv complexity (LZC) estimation and RQA.
	For the purpose of joint LZC estimation, the input matrix was discretized into adjustable number of letters $n_l$ of an alphabet by the given formula. Let $p_{\min}=\min\{p_j|1 \leq j \leq w\}$, $p_{\max}=\max\{p_j|1 \leq j \leq w\}$ and $p_d=p_{max}-p_{min}$ then each element $p_j$ is assigned value $p_j \leftarrow \floor{n_l \frac{p_j-p_{min}}{p_d}}$.
	%
	The joint-LZC therefore stands, in our case, for the complexity of time ordered $n$ dimensional tuples (populations).
	
	In case of RQA, there is a possibility to directly use the spatial data representation \cite{marwan2007generalised}, therefore we did not apply the Takens' embedding theorem and we directly calculated the thresholded similarity matrix from our source data. The RQA features like determinism and divergence were calculated.
	
\begin{figure*}[h]
		\centering
		\includegraphics[trim = 0mm 20mm 0mm 0mm, scale=0.5]{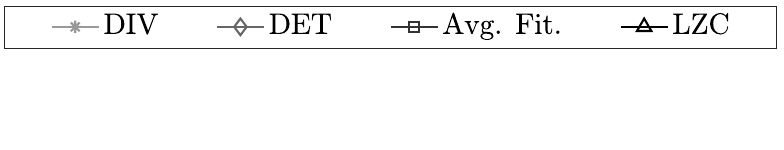}\\
		\subfigure[]{\includegraphics[scale=0.35]{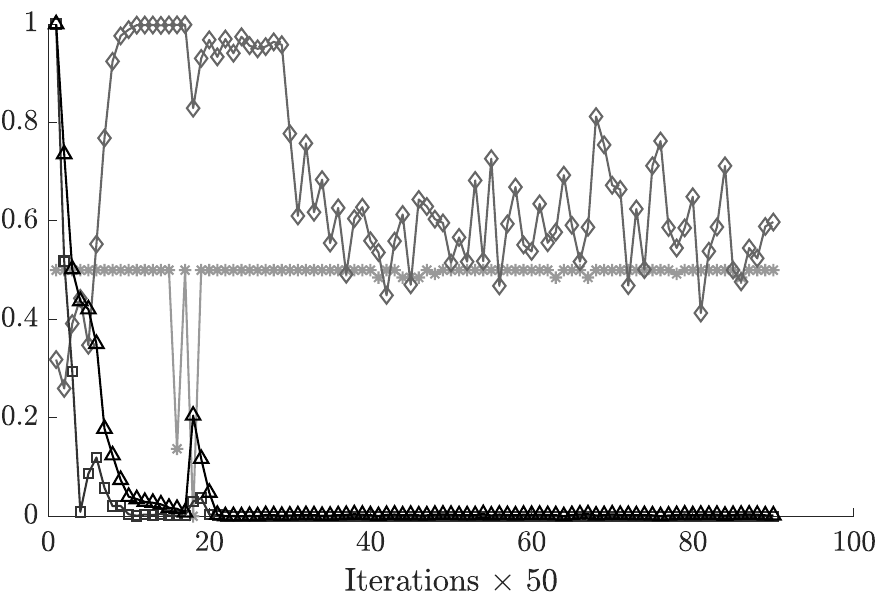}}
		\subfigure[]{\includegraphics[scale=0.35]{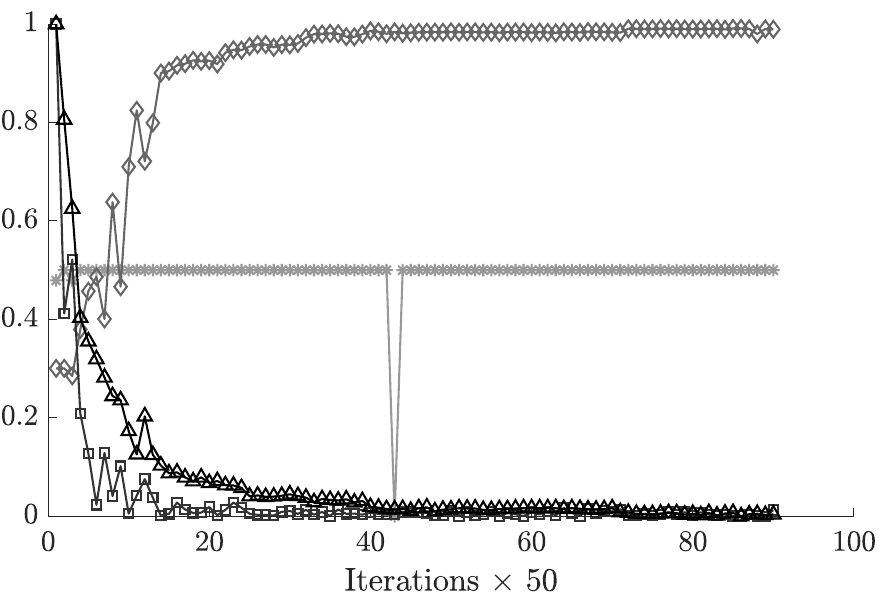}}
		\subfigure[]{\includegraphics[scale=0.35]{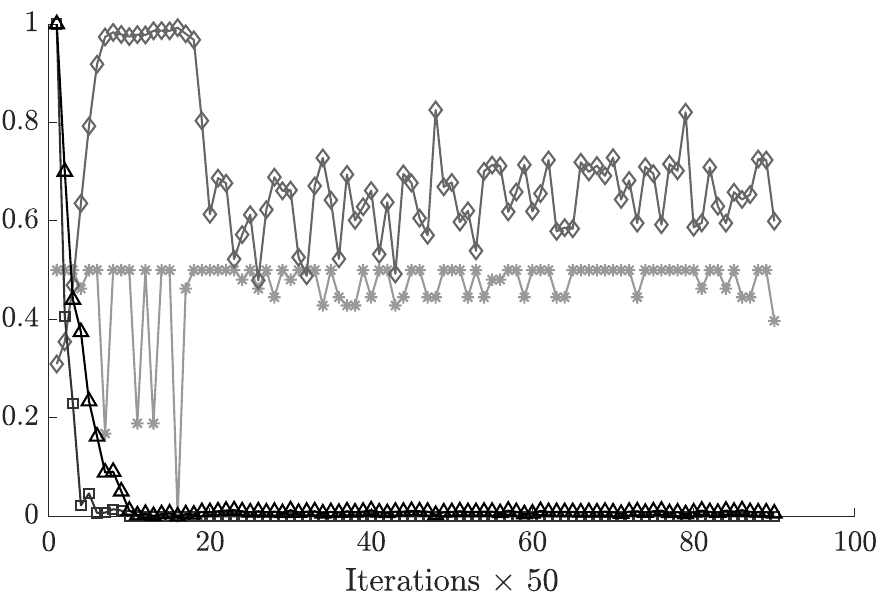}}
		\subfigure[]{\includegraphics[scale=0.35]{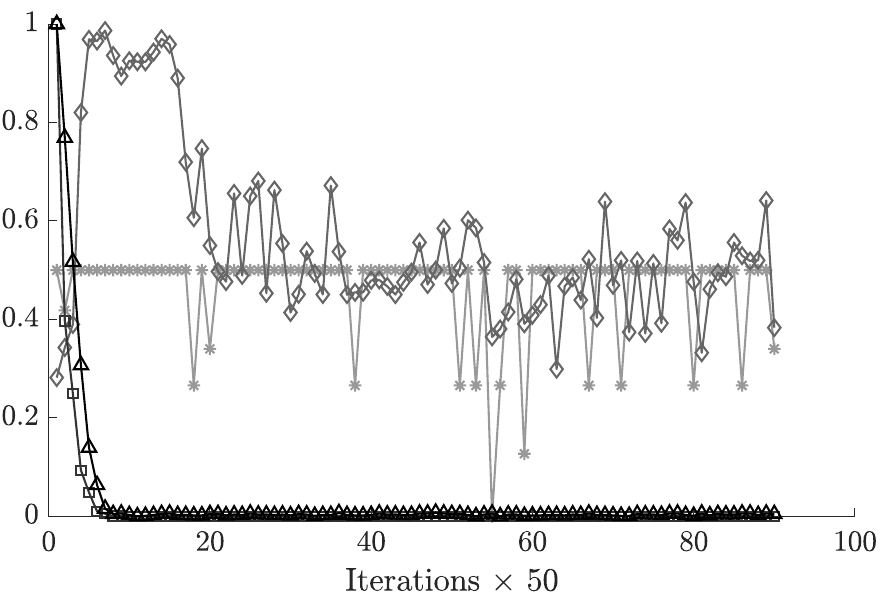}}
		\caption{Progress of the PSO algorithms executed several times with varying populations and window sizes. Horizontal axis represents the migrations while the vertical line holds values of average fitness-function of the population (Avg. Fit.) and obtained indicators. (a)~population size 40, window 20, (b)~population size 70, window 20, (c)~population size 100, window 30, (d)~population size 100, window 40}
		\label{pso_fig}
	\end{figure*}

	
	Based on the obtained visualizations (Fig. \ref{pso_fig}, \ref{de_fig} and \ref{soma_fig}) we are able to confirm the visible differences in cases of PSO and SOMA algorithm. These two optimizations are performing similarities when the population is migrating the same direction. Once the optimum is reached, the similarities decrease. We are not able to confirm the same in case of DE. Due to the randomly performed crossover and additional mutation, this algorithm seems to contain more randomness and evolution-like behavior.
	
	Further examinations calculated the DET, DIV and LZC values during all of the migrations. The statistical difference of these complexity indicators among the converging and non-converging iterations will be examined by ANOVA to confirm the presence of state transitions \cite{larson2008analysis}.
	
	\section{Results}

\begin{figure}[h]
		\centering
		\includegraphics[trim = 0mm 20mm 0mm 0mm, scale=0.5]{prog/legend.pdf}\\
		\subfigure[]{\includegraphics[scale=0.35]{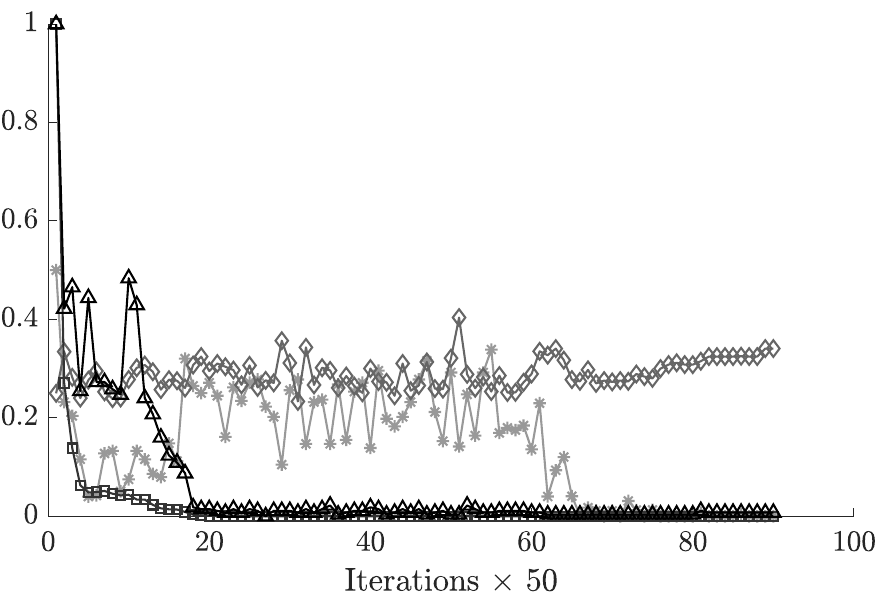}}
		\subfigure[]{\includegraphics[scale=0.35]{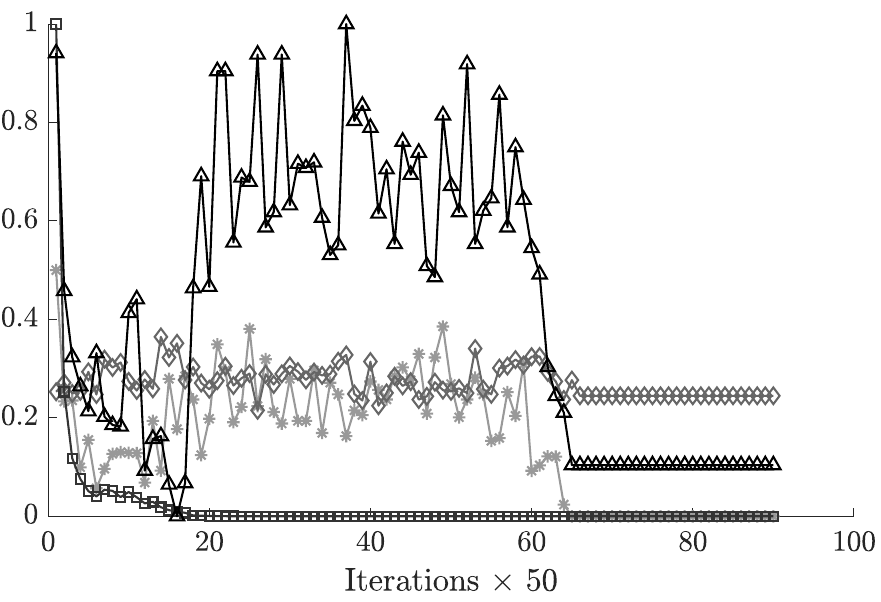}}
		\subfigure[]{\includegraphics[scale=0.35]{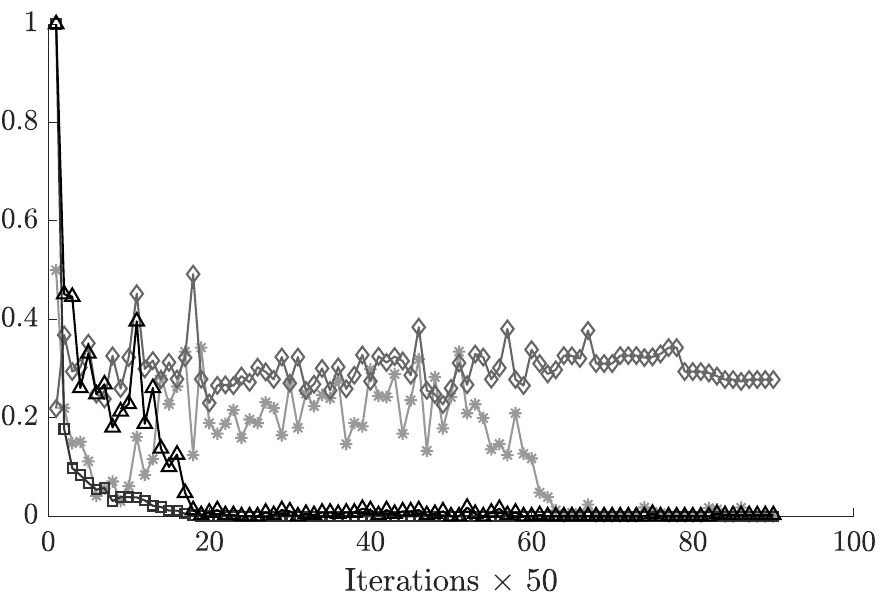}}
		\subfigure[]{\includegraphics[scale=0.35]{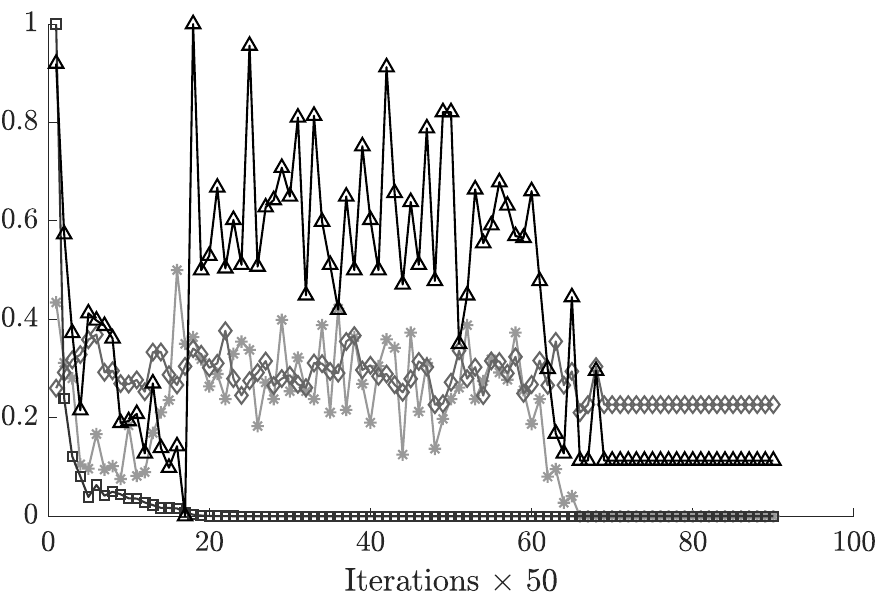}}
		\caption{Progress of the DE algorithms executed several times with varying populations and window sizes. Horizontal axis represents the migrations while the vertical line holds values of average fitness-function of the population (Avg. Fit.) and obtained indicators. (a)~population size 40, window 20, (b)~population size 70, window 20, (c)~population size 70, window 30, (d)~population size 70, window 40}
		\label{de_fig}
	\end{figure}

\begin{figure}[h]
		\centering
		\includegraphics[trim = 0mm 20mm 0mm 0mm, scale=0.5]{prog/legend.pdf}\\
		\subfigure[]{\includegraphics[scale=0.35]{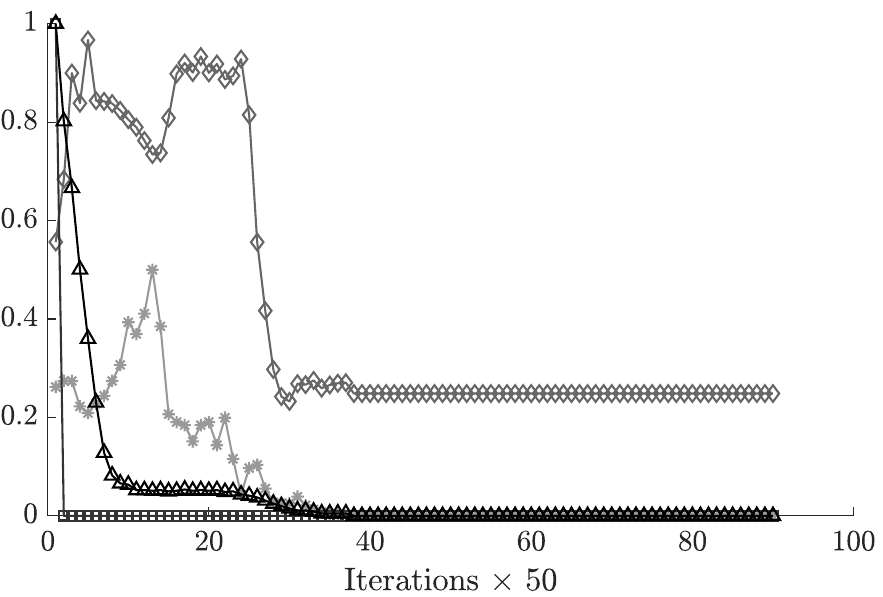}}
		\subfigure[]{\includegraphics[scale=0.35]{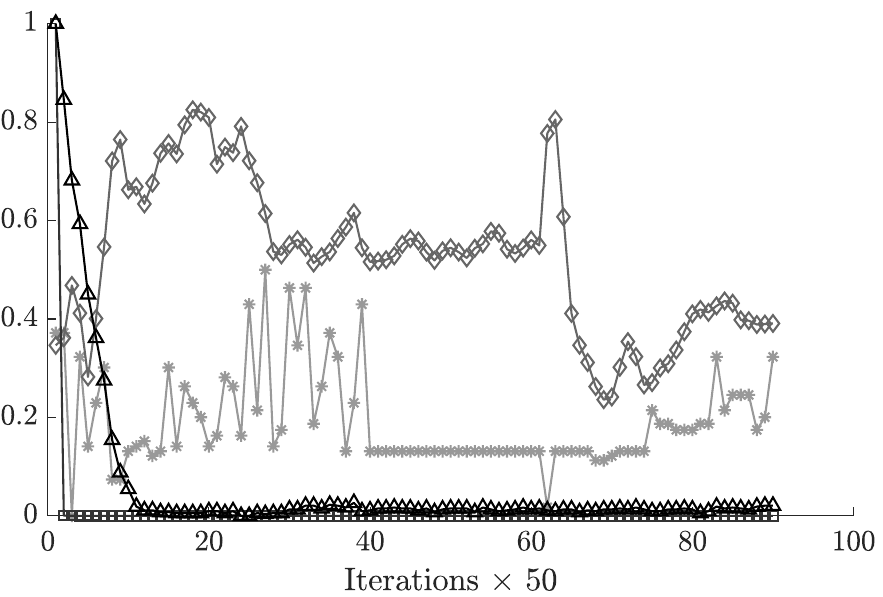}}
		\subfigure[]{\includegraphics[scale=0.35]{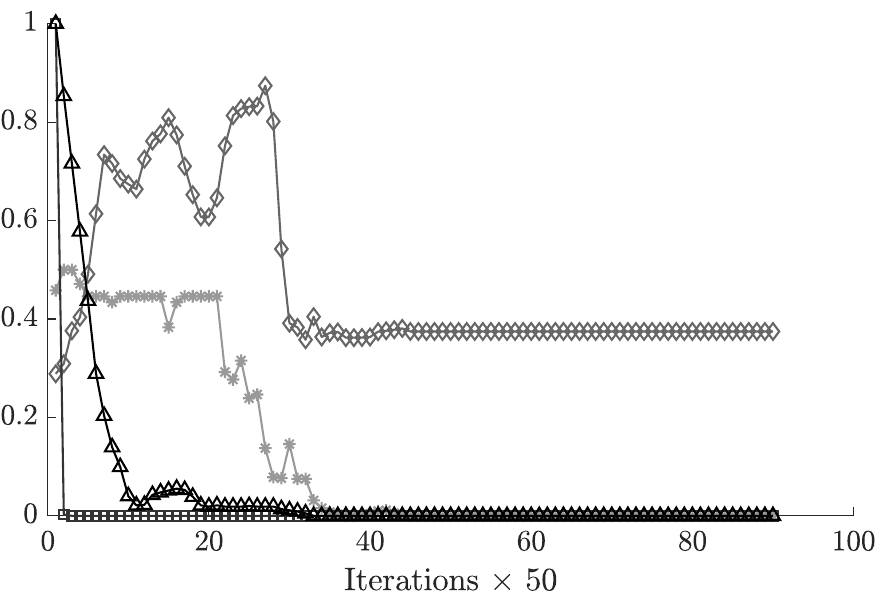}}
		\subfigure[]{\includegraphics[scale=0.35]{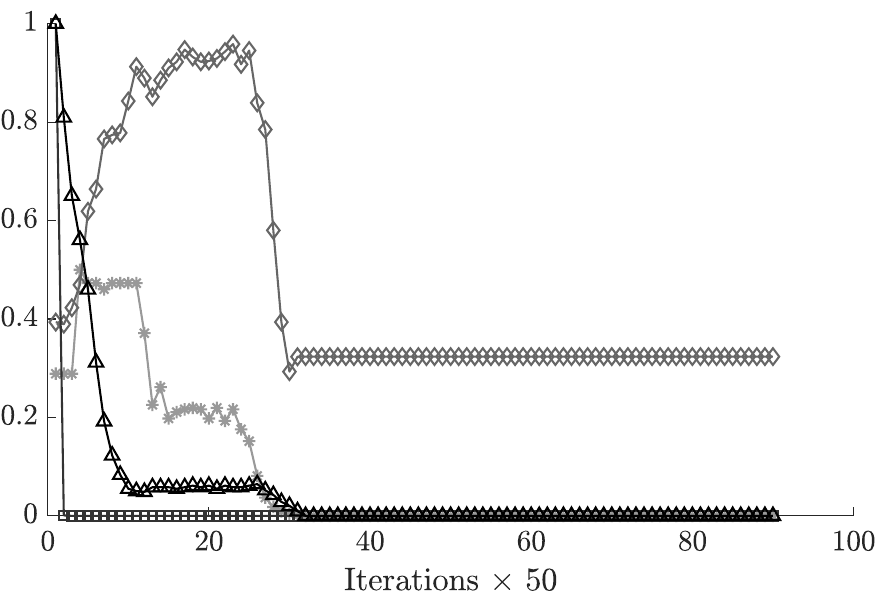}}
		\caption{Progress of the SOMA algorithms executed several times with varying populations and window sizes. Horizontal axis represents the migrations while the vertical line holds values of average fitness-function of the population (Avg. Fit.) and obtained indicators. (a)~population size 40, window 20, (b)~population size 40, window 30, (c)~population size 40, window 40, (d)~population size 100, window 20}
		\label{soma_fig}
	\end{figure}

	 Levels of complexity and the RQA indicators may posses different values based on a given window size as well as the size of the population, therefore we tried several combinations of these parameters (3 per each, therefore nine combinations for each algorithm). Only each tenth value of each time set was plotted in the charts (see Fig. \ref{pso_fig}, \ref{de_fig} and \ref{soma_fig}). The values of fitness-function and LZ complexity were normalized into the range between 0 and 1. The determinism returns such normalized values originally, therefore there was no need for an additional normalization. In case of the divergence, its values were very low ($\times \text{E}10^{-3}$), so it was necessary to multiply them in order to keep the similar visual scale in charts.

	
	\paragraph{Particle swarm optimization.}
	The progress of PSO (Fig \ref{pso_fig}) possess quickly decreasing LZC as the population converges towards an optimum and looses diversity. This behavior is expected as well as some appearing pulses in times when population probably left a local optimum, which was also reflected by an additional converges towards some better solution.
	
	The progress of the population was very much predictable as it was evaluated by DET which possesed values close to 1 when the convergence of the population was the highest. Once a found optimum was reached by the majority of the population, DET dropped and evaluated the population's progress as unpredictable.
	
	Higher values of DIV imply the presence of chaotic behavior in the system. All of the evaluations returned only very small values of this indicator therefore the only small amount of chaos can be confirmed. In the available visual evaluation, the DIV appears to possess the smallest relation to the progress of the algorithm.

	\paragraph{Differential evolution.}
	DE performs elitism during its operation which can be the reason of an absolute flat progress of all its indicators during last iterations. The significant increase of LZC values in some cases remains unclear and can be connected with situation when the population found several optimums of the same quality and the population randomly switched among them (see Fig. \ref{de_fig}).
	The values of DET only evaluate the entire progress of DE as unpredictable almost the same way as the DIV which marked the behavior as chaotic until the found optimum was reached by the population and any other better solution was found.
    
	\paragraph{Self-organizing migration algorithm.}
	The progress of the SOMA algorithm has similarities with both previous algorithms. All indicators are very flat during its last migrations, because particles remains on their positions in cases when better solution was not found. The pertubet following of the leader is similarly reflected by DET as it was in case of PSO, when the behavior of the algorithm was marked as predictable until the majority of the population reached the found optimum. The appearance of the chaos is very low the same way as it was in previous cases (DIV). The LZC as well as the Fitness dropped very quickly because of the nature of SOMA. Each particle performed multiple trials (steps as the path length divided by the step size) and the each population's individual migrated towards its best trial. This is the nature of the algorithm and the reason why it appears as the algorithm with the highest performance in the frame of our experiments.
	
	\subsection{ANOVA testing}
	The DET, DIV and LZC values were split into values obtained in different phases of the optimization. Six groups, marked from 1 to 6, were defined as follows
\begin{itemize}
\item \textbf{1} as progress of PSO algorithm during its converging migrations [10,60]
\item \textbf{2} as progress of DE algorithm during its converging migrations [10,60] 
\item \textbf{3} as progress of SOMA algorithm during its converging migrations [10,60]
\item \textbf{4} as progress of PSO algorithm during its non-converging migrations [300,350]
\item \textbf{5} as progress of DE algorithm during its non-converging migrations [300,350]
\item \textbf{6} as progress of SOMA algorithm during its non-converging migrations [300,350]
\end{itemize}
The presence of statistically significant differences among the means of these groups will confirm the state transitions. Especially we are interested whether the groups of the same algorithms are different and in which indicators.
	
	\begin{figure}[!tbp]
		\includegraphics[scale=0.55]{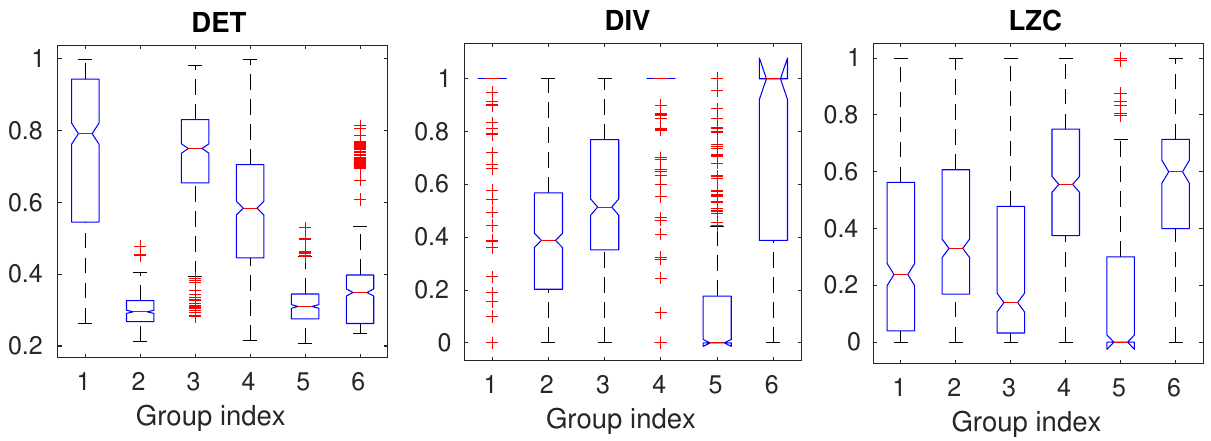}
		\caption{Means with standard deviations obtained by ANOVA testing on six defined groups of data.}
		\label{anova_means}
	\end{figure}
	
	ANOVA testing rejected the null hypothesis that says about similarity of the means across the examined groups of the data (see Fig. \ref{anova_means}). Obtained p-values are 0 for  $\text{ANOVA}_{DET}$ and $\text{ANOVA}_{DIV}$, and $2.657e-94$ for 
	$\text{ANOVA}_{LZC}$.
	The performed additional post-hoc analysis reveled the specific differences among the groups according to their means and it is as follows. The means of Determinism results were able to differentiate the groups 1 and 3 from the rest of the groups, while the means of second group were not significantly different from others (5,6). The separability performance of the means of Divergence were able to significantly exclude the groups 2 and three from the rest while the means of the first group were similar to the fourth group. Both of them differed from the rest significantly. In case of LZC, the groups 1 and 3 are have means significantly different from the rest of the groups while group 2 possesses this difference against all of the groups. 
	
	These results mean that optimization phases are distinguishable by means of this complexity measure. From the above mentioned differences of the means, it is clearly visible that the convergence phases of PSO are separable by the means of Determinism and LZC while in case of Divergence we are not able to distinguish among them. In case of DE, its LZC and Divergence means possessed significant differences between DEs' convergence phases while Determinism was not applicable for this task. And finally the case of SOMA. All of the applied complexity criteria returned significantly different means among the SOMA convergence phases, therefore they are able to be distinguished by these values.

	\section{Discussion}
	
	In contrast to conventional computers, natural systems never stop to function, therefore by simply observing a physical, chemical or living computer we might never know when its completed the task and produced result. This phenomenon was formalized in a framework of inductive Turing machines~\cite{burgin1984inductive} and advanced in structural machines~\cite{burgin2017structural}, however still there is a lack of a definite measure.   Some measures of spatio-temporal dynamics of a computing system are necessary to infer weather consider its current state as representing a final solution or wait longer.
	
	In computer experiments with particle swarm optimization we found that it is possible to detect the convergence of algorithm using RQA and LZ complexity measures. The converging and non-converging iterations of the optimization algorithms are statistically different in the view of applied chaos, complexity and predictability estimating indicators. Typically, the degree of RQA Determinism sharply increases, as if undergoing a phase transition, when fitness approaches its maximum. Dynamics of LZ complexity follows, in general, the level of fitness. These results are well in line, and somewhat complement, our previous studies on the use of dynamics of compressibility of a system's spatial configurations to detect when the system completed computation~\cite{adamatzky2016using}.
	
	Our findings may lead to the future work which is related to the estimation of the edge of chaos in the swarm-like optimization algorithms. It may be applied in a design of adaptive approaches aiming to control their progress in order to sustain the best possible performance. 
	
	
	\section*{ACKNOWLEDGMENT}
The following grants are acknowledged for the financial support provided for this research: Grant of SGS No. SGS 2018/177, VSB-Technical University of Ostrava and German Research Foundation (DFG projects no. MA 4759/9-1 and MA4759/8).

	\bibliographystyle{splncs_srt}
	
	\bibliography{refs}

\begin{thebibliography}{10}

\bibitem{adamatzky2009hot}
Adamatzky, A.:
\newblock Hot ice computer.
\newblock Physics Letters A \textbf{374}(2) (2009)  264--271

\bibitem{adamatzky2016advances}
Adamatzky, A.:
\newblock Advances in Physarum machines: Sensing and computing with slime
  mould. Volume~21.
\newblock Springer (2016)

\bibitem{adamatzky2016using}
Adamatzky, A., Jones, J.:
\newblock On using compressibility to detect when slime mould completed
  computation.
\newblock Complexity \textbf{21}(5) (2016)  162--175

\bibitem{banks2007review}
Banks, A., Vincent, J., Anyakoha, C.:
\newblock A review of particle swarm optimization. part i: background and
  development.
\newblock Natural Computing \textbf{6}(4) (2007)  467--484

\bibitem{bergstra2012random}
Bergstra, J., Bengio, Y.:
\newblock Random search for hyper-parameter optimization.
\newblock Journal of Machine Learning Research \textbf{13}(Feb) (2012)
  281--305

\bibitem{bertschinger2004real}
Bertschinger, N., Natschl{\"a}ger, T.:
\newblock Real-time computation at the edge of chaos in recurrent neural
  networks.
\newblock Neural computation \textbf{16}(7) (2004)  1413--1436

\bibitem{boedecker2012information}
Boedecker, J., Obst, O., Lizier, J.T., Mayer, N.M., Asada, M.:
\newblock Information processing in echo state networks at the edge of chaos.
\newblock Theory in Biosciences \textbf{131}(3) (2012)  205--213

\bibitem{burgin1984inductive}
Burgin, M.:
\newblock Inductive turing machines with a multiple head and kolmogorov
  algorithms.
\newblock In: Soviet Mathematics Doklady. Volume~29. (1984)  189--193

\bibitem{burgin2017structural}
Burgin, M., Adamatzky, A.:
\newblock Structural machines and slime mould computation.
\newblock International Journal of General Systems (2017)  1--24

\bibitem{costello2017calculating}
Costello, B.D.L., Adamatzky, A.:
\newblock Calculating {V}oronoi diagrams using chemical reactions.
\newblock In: Advances in Unconventional Computing.
\newblock Springer (2017)  167--198

\bibitem{cover2012elements}
Cover, T.M., Thomas, J.A.:
\newblock Elements of information theory.
\newblock John Wiley \& Sons (2012)

\bibitem{crutchfield1988computation}
Crutchfield, J.P., Young, K.:
\newblock Computation at the onset of chaos.
\newblock In: The Santa Fe Institute, Westview, Citeseer (1988)

\bibitem{del2008particle}
Del~Valle, Y., Venayagamoorthy, G.K., Mohagheghi, S., Hernandez, J.C., Harley,
  R.G.:
\newblock Particle swarm optimization: basic concepts, variants and
  applications in power systems.
\newblock IEEE Transactions on evolutionary computation \textbf{12}(2) (2008)
  171--195

\bibitem{detrain2006self}
Detrain, C., Deneubourg, J.L.:
\newblock Self-organized structures in a superorganism: do ants “behave”
  like molecules?
\newblock Physics of life Reviews \textbf{3}(3) (2006)  162--187

\bibitem{kadmon2015transition}
Kadmon, J., Sompolinsky, H.:
\newblock Transition to chaos in random neuronal networks.
\newblock Physical Review X \textbf{5}(4) (2015)  041030

\bibitem{kantz97}
Kantz, H., Schreiber, T.:
\newblock {Nonlinear Time Series Analysis}.
\newblock University Press, Cambridge (1997)

\bibitem{kennedy1995particle}
Kennedy, J., Eberhart, R.C.:
\newblock Particle swarm optimization.
\newblock In: Proc. of the IEEE International Conference on Neural Networks.
\newblock (1995)  1942--1948

\bibitem{koebbe1992use}
Koebbe, M., Mayer-Kress, G.:
\newblock Use of recurrence plots in the analysis of time-series data.
\newblock In: SFI Studies in the Sciences of Complexity. Volume~12.,
  Addison-Wesley Publishing (1992)  361--361

\bibitem{langton1990computation}
Langton, C.G.:
\newblock Computation at the edge of chaos: phase transitions and emergent
  computation.
\newblock Physica D: Nonlinear Phenomena \textbf{42}(1-3) (1990)  12--37

\bibitem{larson2008analysis}
Larson, M.G.:
\newblock Analysis of variance.
\newblock Circulation \textbf{117}(1) (2008)  115--121

\bibitem{lempel1976complexity}
Lempel, A., Ziv, J.:
\newblock On the complexity of finite sequences.
\newblock IEEE Transactions on information theory \textbf{22}(1) (1976)  75--81

\bibitem{marwan2007generalised}
Marwan, N., Kurths, J., Saparin, P.:
\newblock Generalised recurrence plot analysis for spatial data.
\newblock Physics Letters A \textbf{360}(4) (2007)  545--551

\bibitem{marwan2007recurrence}
Marwan, N., Romano, M.C., Thiel, M., Kurths, J.:
\newblock Recurrence plots for the analysis of complex systems.
\newblock Physics Reports \textbf{438}(5) (2007)  237--329

\bibitem{mitchell1993revisiting}
Mitchell, M., Hraber, P., Crutchfield, J.P.:
\newblock Revisiting the edge of chaos: Evolving cellular automata to perform
  computations.
\newblock arXiv preprint adap-org/9303003 (1993)

\bibitem{ohira1998phase}
Ohira, T., Sawatari, R.:
\newblock Phase transition in a computer network traffic model.
\newblock Physical Review E \textbf{58}(1) (1998)  193

\bibitem{packard1980geometry}
Packard, N.H., Crutchfield, J.P., Farmer, J.D., Shaw, R.S.:
\newblock Geometry from a time series.
\newblock Physical review letters \textbf{45}(9) (1980)  712

\bibitem{schinkel2008}
Schinkel, S., Dimigen, O., Marwan, N.:
\newblock Selection of recurrence threshold for signal detection.
\newblock European Physical Journal -- Special Topics \textbf{164}(1) (2008)
  45--53

\bibitem{schut2010model}
Schut, M.C.:
\newblock On model design for simulation of collective intelligence.
\newblock Information Sciences \textbf{180}(1) (2010)  132--155

\bibitem{storn1997differential}
Storn, R., Price, K.:
\newblock Differential evolution--a simple and efficient heuristic for global
  optimization over continuous spaces.
\newblock Journal of global optimization \textbf{11}(4) (1997)  341--359

\bibitem{tomaszek2016performance}
Tomaszek, L., Zelinka, I.:
\newblock On performance improvement of the soma swarm based algorithm and its
  complex network duality.
\newblock In: Evolutionary Computation (CEC), 2016 IEEE Congress on, IEEE
  (2016)  4494--4500

\bibitem{wright2001cyclic}
Wright, A.H., Agapie, A.:
\newblock Cyclic and chaotic behavior in genetic algorithms.
\newblock In: Proceedings of the 3rd Annual Conference on Genetic and
  Evolutionary Computation, Morgan Kaufmann Publishers Inc. (2001)  718--724

\bibitem{zbilut1992embeddings}
Zbilut, J.P., Webber, C.L.:
\newblock Embeddings and delays as derived from quantification of recurrence
  plots.
\newblock Physics letters A \textbf{171}(3-4) (1992)  199--203

\bibitem{zbilut2002recurrence}
Zbilut, J.P., Zaldivar-Comenges, J.M., Strozzi, F.:
\newblock Recurrence quantification based liapunov exponents for monitoring
  divergence in experimental data.
\newblock Physics Letters A \textbf{297}(3) (2002)  173--181

\bibitem{zelinka2004soma}
Zelinka, I.:
\newblock Soma—self-organizing migrating algorithm.
\newblock In: New optimization techniques in engineering.
\newblock Springer (2004)  167--217

\bibitem{zelinka2017novel}
Zelinka, I., Tomaszek, L., Vasant, P., Dao, T.T., Hoang, D.V.:
\newblock A novel approach on evolutionary dynamics analysis--a progress
  report.
\newblock Journal of Computational Science (2017)

\bibitem{zenil2017algorithmic}
Zenil, H., Gauvrit, N.:
\newblock Algorithmic cognition and the computational nature of the mind.
\newblock Encyclopedia of Complexity and Systems Science (2017)  1--9

\bibitem{zozor2005lempel}
Zozor, S., Ravier, P., Buttelli, O.:
\newblock On lempel--ziv complexity for multidimensional data analysis.
\newblock Physica A: Statistical Mechanics and its Applications \textbf{345}(1)
  (2005)  285--302

\end{thebibliography}


\end{document}